\documentstyle[12pt]{article}
\begin{document}

\centerline{\bf Stability of the E2/M1 ratio}
\centerline{\bf at the T-matrix pole}
\vskip .3cm
\centerline{Ron L. Workman and Richard A. Arndt}
\centerline{Department of Physics, Virginia Tech, Blacksburg, VA 24061 }

\begin{abstract}

We show how the stability of the E2/M1 ratio,
evaluated at the T-matrix pole, can be understood given
a much wider variation at the K-matrix pole.

\end{abstract}

\vskip .5cm

Interest in the E2/M1 ratio has recently increased, due largely to
new and precise pion photoproduction experiments spanning
the first resonance region\cite{expts}. Several novel theoretical
works have also attempted to relate this quantity to predictions given
by quark models\cite{Buchmann,Aznauryan}. While this ratio of electric
quadrupole (E2) and magnetic dipole (M1) amplitudes is usually evaluated
at the K-matrix pole, the Mainz group\cite{HDT} has shown that it
is particularly stable when evaluated at the T-matrix pole position.
Models which gave K-matrix ratios differing by factors of two or three
were shown to have very similar T-matrix ratios. This stability was
subsequently verified by both the VPI and RPI groups\cite{vpi_pole,pc}.
In the following, we show how these features can be understood within
a simple parameterization scheme.

At the T-matrix pole, the E2 and M1 residues are complex,
resulting in a complex ratio. This differs from the K-matrix ratio
(Im$E^{3/2}_{1+}$/Im$M^{3/2}_{1+}$, at the point where
Re$M_{1+}^{3/2} \to 0$) which gives a purely real
number. These residues are determined either via the ``speed plot''
method\cite{HDT} or by continuing the parameters of a model into
the complex energy plane\cite{vpi_pole}. Quantitative results vary
slightly, depending on details of the extrapolation
procedure\cite{Wilbois}. In the present study we have continued our
fits to the pole. Qualitatively similar results should follow from
speed plot determinations.

We have fitted the pion photoproduction database
using multipoles of the form
\begin{equation}
M = \alpha ( 1 + i T_{\pi N} ) + \beta T_{\pi N} ,
\end{equation}
where $\alpha$ includes the projected Born terms and both $\alpha$
and $\beta$ contain kinematic factors required for correct threshold
behavior. The $\pi N$ T-matrix
($T_{\pi N}$) enters this parameterization in a direct way and
provides most of the required energy dependence.
For energies of interest, the T-matrix is essentially elastic, and
thus the first term is proportional to $\cos \delta e^{i\delta}$,
$\delta$ being the associated $\pi N$ phase shift. A term of this
form arises naturally in many unitarization schemes, dynamical models,
and solutions of the associated
dispersion relations\cite{CGLN,Omnes,DMW,CD}. The second term, proportional
to $T_{\pi N}$ accounts for the energy dependence of the dominant M1
amplitude and gives a non-zero E2/M1 ratio at the K-matrix pole (the
first term, being proportional to $\cos \delta$, vanishes at this
point). Phenomenological polynomials in energy were included in both
$\alpha$ and $\beta$ in order to account for missing energy dependence.
However, in fitting
the restricted energy range (250--450 MeV) covering the
delta resonance, it was found that only a constant factor was required
for $\beta$.
In addition, since the Born terms are slowly varying in this region,
our extrapolations to the T-matrix pole required only a simple linear
function of energy for $\alpha$.

Results for the E2/M1 ratio, (at both the K- and
T-matrix pole) using the above form, have closely matched those
derived using either effective Lagrangian or dispersion-relation
approaches, when similar databases have been fitted\cite{DTS}.
This suggests that our results should (at least qualitatively)
agree with those found using other methods.

In Table I, we compare fits to two different databases\cite{rpi_vpi}
which result
in very different E2/M1 ratios at the K-matrix pole. Values for the
E2 and M1 residues at the T-matrix pole are also given. These have
been split to show the contributions from the first and second terms
in Eq.~(1). This comparison is particularly revealing, as the first
term vanishes at the K-matrix pole but does not vanish at the
T-matrix pole. In fact, the term proportional to $(1+iT_{\pi N})$
gives the dominant contribution to E2 at the T-matrix pole. As expected,
the second term, proportional to $T_{\pi N}$, dominates M1. This
immediately explains why the T-matrix pole ratio is so stable. The
dominant contribution to E2 is controlled mainly by the Born terms,
which are essentially fixed in any model\cite{BRNX}.
It is also instructive to compare the variation
of the $\beta$ term at both the K- and T-matrix poles. In the two
fits we have presented, the K-matrix pole ratio varies by a factor
of five. The same level of variation is seen in $\beta$
at the T-matrix pole.
In fact, within this simple parameterization, the ratio
$\beta_{\rm E2} / \beta_{\rm M1}$ at the T-matrix
pole is equivalent to the full E2/M1 ratio at the K-matrix pole.

From the above discussion it should be clear that the T-matrix pole
ratio will remain finite even if the K-matrix ratio is zero. Thus,
as has been suggested previously\cite{Wilbois}, these two ratios
probe different parts of the photoproduction amplitude. Within our
parameterization scheme, there is a simple relation
between the K- and T-matrix ratios
\begin{equation}
E2/M1 |_{\rm T-matrix} \approx E2/M1 |_{\rm K-matrix} +
  i\alpha_{\rm E2}/ \beta_{\rm M1} |_{W_{\rm pole} }  ,
\end{equation}
if one neglects $\alpha_{\rm M1 } / \beta_{\rm M1 }$, which from Table~I
is a good approximation. Since $\alpha$ and $\beta$ are functions of
energy, they become complex at the T-matrix pole ($W_{\rm pole}$) and
contribute to both the real and imaginary
parts of the T-matrix ratio\cite{Resnick}.
From Table~I, we see that the term $i\alpha_{\rm E2}/ \beta_{\rm M1}$
is about 0.05 in modulus, and is predominantly imaginary. This gives
a rough way to see why the imaginary part of the T-matrix pole ratio
remains more stable than the real part, as the K-matrix ratio goes
to zero.

In summary, we have given a simple way to understand the relative
features of the E2/M1 ratio at the K- and T-matrix poles. We hope
this can be of some pedagogical value. The more difficult
question of a separation into bare/dressed or resonant/background
contributions must still be addressed if one wishes to compare with
models.

We thank L. Tiator and R.M. Davidson for many helpful discussions
regarding the extraction of the E2/M1 ratio at the T-matrix pole.
This work was supported in part by a U.S. Department of Energy Grant No.
DE-FG02-97ER41038.

\newpage
\begin{table}
\caption{Comparison of K-matrix and T-matrix results for the E2/M1 ratio.
Fits A and B result in E2/M1 ratios, at the K-matrix pole, of
$-1.93\%$ and $-0.39\%$ respectively. At the T-matrix pole, Fits A and B
have E2/M1 ratios of (modulus, phase) $(0.066,-127^\circ )$ and
$(0.049,-100^\circ )$ respectively.  Here $\alpha$ and $\beta$ refer to
the two terms in Eq.~(1).}
\label{table1}
\begin{center}
\begin{tabular}{|c|c|c|c|}\hline
 & \multicolumn{3}{c|}{ E2 } \\ \cline{2-4}
  &$\alpha$-term  & $\beta$-term & Total \\ \hline
Fit A  & $(1.12,\; -143^\circ )$ & $(0.42,\; 157^\circ)$ &
$(1.38,\; -158^\circ )$ \\
     &     &     &    \\ \hline
Fit B & $(0.98,\; -126^\circ )$  &  $(0.08,\; 157^\circ )$ &
$(1.01,\; -131^\circ )$ \\
     &    &    &    \\ \hline
 & \multicolumn{3}{c|}{ M1 } \\ \cline{2-4}
  &$\alpha$-term  & $\beta$-term & Total \\ \hline
Fit A  & $(3.2,\; -136^\circ )$ &
$(21.9, \; -23^\circ )$ & $(20.9, \; -31^\circ )$ \\
     &    &    &    \\ \hline
Fit B & $(3.2, \; -136^\circ )$  &
  $(21.7, \; -23^\circ )$ & $(20.7,\; -31^\circ )$ \\
     &    &    &    \\ \hline
\end{tabular}
\end{center}
\end{table}


\begin{thebibliography}{99}

\bibitem{expts}
G.Blanpied {\it et al.}, Phys. Rev. Lett {\bf 76}, 1023 (1997);
R. Beck {\it et al.}, Phys. Rev. Lett. {\bf 79}, 4337 (1997).

\bibitem{Buchmann}
A.J. Buchmann, E. Hern\'andez, and A. Faessler,
Phys. Rev. C {\bf 55}, 448 (1997).

\bibitem{Aznauryan}
I.G. Aznauryan,
Phys. Rev. D {\bf 57}, 2727 (1998);
T.~Sato and T.-S.H. Lee,
Phys. Rev. C {\bf 54}, 2660 (1996).
Both of these works extract values for the M1 amplitude
in approximate agreement with quark model predictions.
However, Aznauryan considers her result to be dressed,
while the Sato and Lee value is clearly for a bare
amplitude.

\bibitem{HDT}
O. Hanstein, D. Drechsel, and L. Tiator,
Nucl. Phys. {\bf A632}, 561 (1998);
Phys. Lett. B {\bf 385}, 45 (1996);
L.~Tiator, D.~Drechsel, and O.~Hanstein,
{\it Proceedings of the 4th CEBAF/INT Workshop
on N$^*$ Physics}, ed. by T.-S.H. Lee and W. Roberts
(World Scientific, Singapore, 1997), p. 296.

\bibitem{vpi_pole}
R.A. Arndt, I.I. Strakovsky, and R.L. Workman,
Phys. Rev. C {\bf 56}, 577 (1997).

\bibitem{pc}
R.M. Davidson and N.C. Mukhopadhyay, private communications.

\bibitem{Wilbois}
Th. Wilbois, P. Wilhelm, and H. Arenh\"ovel,
Phys. Rev. C {\bf 57}, 295 (1998);
P. Wilhelm, Th. Wilbois, and H. Arenh\"ovel,
Phys. Rev. C {\bf 54}, 1423 (1996).

\bibitem{CGLN}
G.F. Chew, M.L. Goldberger, F.E. Low, and Y. Nambu,
Phys. Rev. {\bf 106}, 1345 (1957).

\bibitem{Omnes}
R. Omn\`es,
Il Nuovo Cimento {\bf 8}, 316 (1958);
N.~Muskhelishvili, {\it Singular Integral Equations}
(Nordhoff, Groningen, 1953).

\bibitem{DMW}
R.M. Davidson, N.C. Mukhopadhyay, and R.S. Wittman,
Phys. Rev. D {\bf 43}, 71 (1991).

\bibitem{CD}
See, for example, S.N. Yang,
J. Phys. G {\bf 11}, L205 (1985); S. Nozawa, B. Blankleider, and
T.-S.H. Lee, Nucl. Phys. {\bf A513}, 459 (1990); P. Christillin
and G. Dillon, J. Phys. G {\bf 22}, 1773 (1996).

\bibitem{DTS}
R.M. Davidson, L. Tiator and A. Sandorfi, private communications. See
also Refs.\cite{expts,HDT}.

\bibitem{rpi_vpi}
These databases were constructed in conjunction with R.M. Davidson,
N.C. Mukhopadhyay and M.S. Pierce in order to gauge the model- and
database-dependence of the E2/M1 ratio. They differ mainly in the
inclusion/exclusion of certain precise wide-angle Bonn cross section
measurements [ H.~Genzel {\it et al.}, Z. Physik {\bf 268}, 43 (1974),
G.~Fischer {\it et al.}, Z. Physik {\bf 245}, 225 (1971) ].
R.M. Davidson {\it et al.}, submitted for publication.

\bibitem{BRNX}
The contribution from Born terms times $\cos \delta e^{i\delta }$ can
be compared to the phenomenological multipole amplitudes using the
SAID program ( Telnet: said.phys.vt.edu, userid: said ).

\bibitem{Resnick}
There would be a sharper
correspondence between the real part of the T-matrix
pole ratio and the (real) K-matrix value if $\alpha$ and $\beta$ were
energy independent. See L. Resnick, Phys. Rev. D {\bf 2}, 1975 (1970)
for an interesting and related study exploring the consequences of
energy-independent Born terms in the Muskhelishvili-Omn\`es formalism.

\end{thebibliography}
\end{document}